\documentclass[12pt,preprint]{aastex}
\bibpunct{(}{)}{;}{a}{}{,}
\let\tmpclearpage\clearpage
\let\clearpage\relax

\newcommand{\figureone}[3]{%
\begin{figure}[tbp]
\begin{center}
\epsscale{0.48}\plotone{#1}
\caption{#3}
\label{#2}
\end{center}
\end{figure}
}

\newcommand{\figuretwo}[3]{%
\begin{figure*}[tbp]
\begin{center}
\epsscale{0.96}\plotone{#1}
\caption{#3}
\label{#2}
\end{center}
\end{figure*}
}

\newcommand\CaIIH{\mbox{Ca\,\textsc{ii}\,H}}
\newcommand\CaIIK{\mbox{Ca\,\textsc{ii}\,K}}
\newcommand\CaIIHK{\mbox{Ca\,\textsc{ii}\,H~\&~K}}
\newcommand\FeI{\mbox{Fe\,\textsc{i}}} 

\newcommand\Mm{\ensuremath{\mathrm{Mm}}}

\newcommand\nm{\ensuremath{\mathrm{nm}}}
\newcommand\picom{\ensuremath{\mathrm{pm}}}
\newcommand\kms{\ensuremath{\mathrm{km}/\mathrm{s}}}
\newcommand\mxsqcm{\ensuremath{\mathrm{Mx}/\mathrm{cm}^2}}
\newcommand\s{\ensuremath{\mathrm{s}}}

\begin{document}

\title{Hinode Observations of Magnetic Elements in Internetwork Areas}
\shorttitle{Magnetic Elements in Internetwork Areas}
\author{A.G.~de~Wijn and B.W.~Lites}
\email{dwijn@ucar.edu}
\affil{High Altitude Observatory, National Center for Atmospheric Research\altaffilmark{1}, P.O. Box 3000, Boulder, CO 80307, USA}
\and
\author{T.E.~Berger, Z.A.~Frank and T.D.~Tarbell}
\affil{Lockheed Martin Solar and Astrophysics Laboratory, Palo Alto, CA 94304, USA}
\and
\author{R.~Ishikawa}
\affil{Department of Astronomy, University of Tokyo, Hongo, Bunkyo-ku, Tokyo 113-0033, Japan}
\shortauthors{De Wijn et al.}
\altaffiltext{1}{The National Center for Atmospheric Research is sponsored by the National Science Foundation.}

\begin{abstract}
We use sequences of images and magnetograms from Hinode to study magnetic elements in internetwork parts of the quiet solar photosphere.
Visual inspection shows the existence of many long-lived (several hours) structures that interact frequently, and may migrate over distances $\sim7~\Mm$ over a period of a few hours.
About a fifth of the elements have an associated bright point in G-band or \CaIIH\ intensity.
We apply a hysteresis-based algorithm to identify elements.
The algorithm is able to track elements for about $10~\min$ on average.
Elements intermittently drop below the detection limit, though the associated flux apparently persists and often reappears some time later.
We infer proper motions of elements from their successive positions,
and find that they obey a Gaussian distribution with an rms of $1.57\pm0.08~\kms$.
The apparent flows indicate a bias of about $0.2~\kms$ toward the network boundary.
Elements of negative polarity show a higher bias than elements of positive polarity, perhaps as a result of to the dominant positive polarity of the network in the field of view, or because of increased mobility due to their smaller size.
A preference for motions in $X$ is likely explained by higher supergranular flow in that direction.
We search for emerging bipoles by grouping elements of opposite polarity that appear close together in space and time.
We find no evidence supporting Joy's law at arcsecond scales.
\end{abstract}


\section{Introduction}\label{sec:introduction}

Magnetic elements have been extensively studied in network that partially outlines the boundaries of supergranular cells.
They were first observed as ``magnetic knots''
	\citep{1968SoPh....4..142B} 
and as ``filigree''
	\citep{1973SoPh...33..281D}, 
before being resolved into strings of adjacent bright points by
	\cite{1974SoPh...38...43M}. 
	\cite{1977SoPh...52..249M} 
and
	\cite{1981SoPh...69....9W} 
showed that faculae, filigree, and bright points in wide-band \CaIIH\ filtergrams are manifestations of the same phenomenon.
	\cite{1983SoPh...85..113M}  
introduced the name ``network bright point'', and subsequently initiated extensive studies of magnetic elements as G-band bright points
	\citep{1984SoPh...94...33M}. 
Studies of bright points using high-resolution imaging 
	\citep[e.g.,][]{1995ApJ...454..531B, 
	1998ApJ...506..439B, 
	1998ApJ...495..973B, 
	2004A&A...428..613B, 
	1996ApJ...463..365B, 
	2001ApJ...553..449B, 
	2005A&A...435..327R, 
	2007A&A...466.1131R, 
	2007A&A...472..911I, 
	2007ApJ...661.1272B} 
have since established that network bright points are manifestations of small, kilogauss magnetic elements that form the magnetic network
	\citep{1968SoPh....5..442C, 
	1969SoPh...10..294L, 
	1972SoPh...22..402H, 
	1972SoPh...27..330F, 
	1973SoPh...32...41S}. 

Magnetic field in internetwork have been largely ignored until recently.
However, it is currently being studied vigorously
	\citep[e.g.,][]{2003ApJ...582L..55D, 
	2003ApJ...597L.177S, 
	2004ApJ...611.1139S, 
	2004ApJ...613..600L, 
	2004ApJ...616..587S, 
	2004Natur.430..326T, 
	2004ApJ...614L..89M, 
	2005A&A...436L..27K, 
	2006ApJ...636..496C, 
	2007A&A...476L..33R, 
	2007ApJ...657.1150S, 
	2007ApJ...659L.177H, 
	2007ApJ...670L..61O, 
	2008ApJ...672.1237L}. 
Many of these studies focus on determining the strength and distribution of flux.
While there is some disagreement between results, it seems that field is ubiquitously present in internetwork at small scales.

Concentrations of flux that are sufficiently strong may form internetwork bright points.
Their existence was already noted by
	\cite{1983SoPh...85..113M}. 
Few recent studies have analyzed these internetwork magnetic elements.
	\cite{2004ApJ...609L..91S} 
measured internetwork bright point density and lifetime,
	\cite{2005A&A...441.1183D} 
reported that internetwork bright points trace locations of flux that may persist for periods of hours,
	\cite{2007A&A...462..303T} 
analyzed morphology, dynamics, and evolution of bright points in \CaIIK\ in quiet sun, and
	\cite{2007A&A...475.1101S} 
searched for photospheric foot points of transition region loops in quiet sun.

Magnetic elements were first modeled as ``flux tubes'' by
	\cite{1976SoPh...50..269S}. 
Over the years, models grew increasingly complex
	\citep[e.g.,][]{1988A&A...202..275K, 
	1990A&A...233..583K, 
	1992A&A...262L..29S, 
	1994A&A...285..648G, 
	1998A&A...337..928G, 
	1998ApJ...495..468S, 
	2005A&A...430..691S}. 
The explanation of photospheric brightness enhancement of faculae due to hot walls proposed by
	\cite{1981SoPh...70..207S} 
was verified by MHD simulations by
	\cite{2004ApJ...607L..59K} 
and
	\cite{2004ApJ...610L.137C}. 
On disk, bright points are formed as a result of radiation escaping from deeper, hotter layers due to the fluxtube Wilson depression.

Some authors have noted that magnetic fields in internetwork areas appear to outline cells on mesogranular scales
	\citep[e.g.,][]{2003ApJ...582L..55D, 
	2005A&A...441.1183D, 
	2007A&A...462..303T, 
	2008ApJ...672.1237L}, 
while
	\citet{1998ApJ...495..973B} 
observed ``voids'' in active network.
In addition, recent simulations indicate that field concentrates on boundaries of mesogranular cells
	\citep{2006ApJ...642.1246S}. 
One would expect such a pattern to be set by granular motions, similar to supergranular flows that eventually advect internetwork field into network
	\citep[e.g.,][]{2000SoPh..197...21L}. 
Perhaps magnetic elements form these patterns as a result of flows associated with ``trees of fragmenting granules''
	\citep{2004A&A...419..757R}, 
which were previously linked to mesogranules by
	\citet{2003A&A...409..299R}. 
Flux is expunged by the sideways expansion of granular cells, and is collected in the downflows in intergranular lanes.
In a ``tree of fragmenting granules'', these flows would be expected to drive flux not only to the edges of individual granules, but also to the edges of the tree.

In this paper, we present a study of the dynamics of magnetic elements in internetwork parts of the solar photosphere.
This study is motivated by its relevance to the operation of a turbulent granular dynamo, the nature of quiet-sun magnetism, the generation of MHD waves that may propagate into the transition region and corona, and the coupling of internetwork field to the magnetic network.
First, examples of magnetic elements are discussed in the context of fluxtube dynamics and lifetime (Sect.~\ref{sec:visual}).
Magnetic elements are compared with bright points in G-band and \CaIIH\ intensity in Sect.~\ref{sec:comparebp}.
A feature-tracking algorithm is applied in order to analyze the lifetime (Sect.~\ref{sec:lifetime}) and the dynamics (Sects.~\ref{sec:velocities} and~\ref{sec:direction}) of magnetic elements.
Finally, a search for emerging bipoles is presented in Sect.~\ref{sec:emerging}.

\section{Observations and Data Reduction}\label{sec:observations}

\clearpage
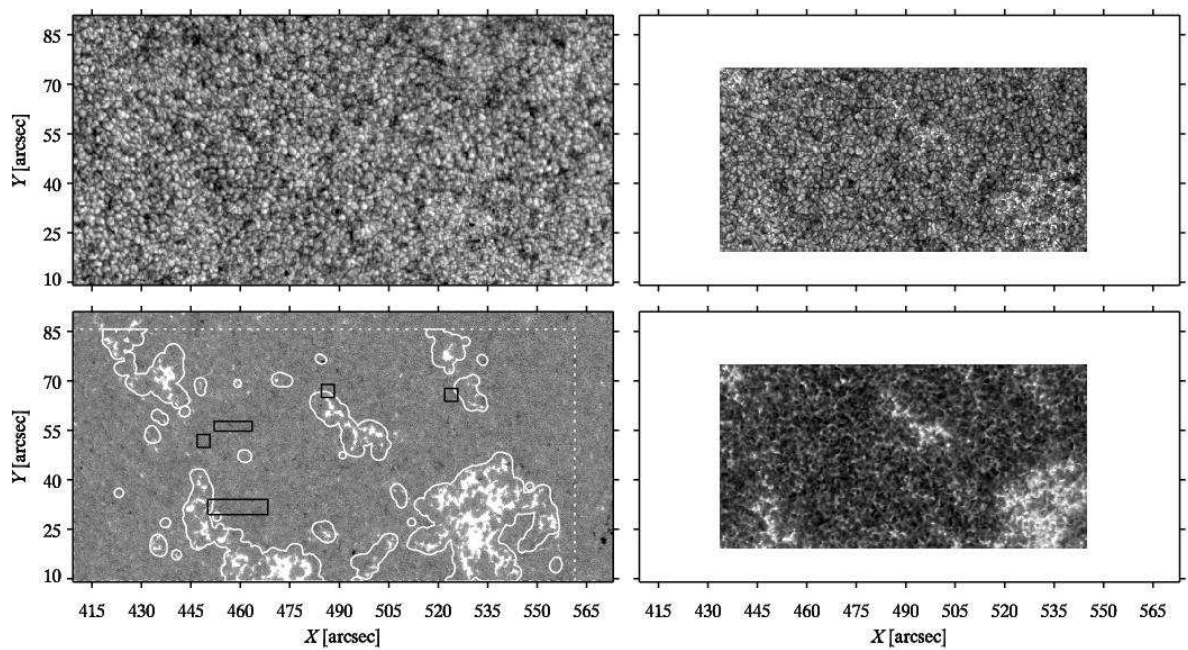
\figuretwo{f1}{fig:sample}{A set of sample images from the sequences.
Clockwise from top left: \FeI\ intensity, G-band intensity, \CaIIH\ intensity, and \FeI\ magnetogram scaled between $-500$ (black) and $+500~\mxsqcm$ (white).
Solid white contours drawn on the magnetogram outline the network.
The area contained by the dotted white lines is not affected by incomplete sampling due to image motion.
Regions of interest examined in Figs.~\ref{fig:slicesample1}, \ref{fig:slicesample2}, and~\ref{fig:globsamples} are outlined by black lines.}
\clearpage

We use an image sequence of a quiet area recorded by \emph{Hinode}
	\citep{2007SoPh..243....3K} 
using the Solar Optical Telescope
	\citep{Tsuneta2007,Suematsu2007,Ichimoto2007,2007SoPh..tmp..154S,2007SoPh..243...87M}
from 00:18 to 06:00~UT on March~30, 2007.
\emph{Hinode} was programmed to observe quiet sun near a small area of weak plage.
The center of the field of view was at $(490\arcsec,50\arcsec)$ at the beginning of the sequence, and approximately followed solar rotation during the sequence.
Because $\mu$ ranges from $0.91$ to $0.75$ over the field of view, care must be taken to correct measurements of position and velocity for foreshortening.
Image sequences were recorded in the G band and in the \CaIIH\ line using the Broadband Filter Imager.
The latter have chromospheric contributions, but mostly sample the upper photosphere due to the broad filter bandwidth of $0.3~\nm$.
The Narrowband Filter Imager was used to record Stokes I and V in the photospheric \FeI\ line at $630.2~\nm$ at an offset of $-12~\picom$ from line center.
The spectral resolution of the filter is $9~\picom$ at this wavelength.
The Stokes V signal is sensitive to Dopplershift as a result of flows along the line of sight.
It is impossible to compensate for this effect due to the single line position used in the observations.
The frames consist of $1024\times512$ square pixels.
Each pixel in a frame corresponds to $2\times2$ pixels on the CCD, summed to sacrifice resolution for increased cadence and field of view at constant telemetry.
The G-band and \CaIIH\ images have a pixel scale of $0.11\arcsec$ and a field of view of $112\arcsec\times56\arcsec$.
The \FeI\ images have a pixel scale of $0.16\arcsec$, resulting in a larger field of view of $164\arcsec\times82\arcsec$.
A total of 582~frames were recorded in each passband at regular cadence of $35.14~\s$.

The G-band, \CaIIH\ and \FeI\ filtergrams were corrected for dark current and were flatfielded using the SolarSoft procedure \texttt{fg\_prep}.
For the \FeI\ data, a custom flatfield was derived from \FeI\ images recorded between 00:18 and 09:25~UT on March~30, 2007.
This comprises the sequence described above, and a slightly shorter, but otherwise identical sequence of 324~frames of the same area.
A synoptic observation separates the two sequences.
The \FeI\ flatfield is created by averaging the raw data of both sequences.
Visual inspection of the resulting flatfield shows little remaining signal of solar origin, and clearly shows the CCD fringe pattern.
We apply a simple calibration of the Stokes V images to produce magnetograms following
	\cite{chae2007}. 

We carefully align the frames using Fourier cross-correlation techniques.
The sequences of magnetograms, G-band images, and \CaIIH\ images are first aligned separately.
The displacements computed from magnetograms are also applied to associated \FeI\ filtergrams.
Next, the G-band and \CaIIH\ sequences are aligned to the \FeI\ intensity and unsigned magnetogram sequences, respectively.
Figure~\ref{fig:sample} shows a sample \FeI\ filtergram, the associated magnetogram, G-band image, and \CaIIH\ image.

A mask of network areas is created by taking a suitable threshold of the average magnetogram, smoothed to remove structures with sizes below $2\arcsec$.
The network mask is shown in the bottom left panel of Fig.~\ref{fig:sample}.

We identify areas of significant magnetic flux along the line of sight using a hysteresis-based algorithm.
In the remainder of this paper, ``flux'' refers to flux along the line of sight, unless otherwise indicated.
Such an algorithm searches for features in data using a low threshold, but accepts a feature only if it is also detected using a high threshold
  \cite[e.g.,][]{2007ApJ...666..576D}. 
The low threshold must be chosen such that features close to the noise level are still identified, while the high threshold minimizes the number of false positives.
We begin by producing a running average of three frames of the sequence of magnetograms.
Then, each frame is convolved with a round kernel with a diameter of 5~pixels, and also with one with a 11-pixel diameter.
At this point, we split the analysis between positive and negative flux by the sign of the frame convolved with the 11-pixel kernel.
We apply a suitable threshold to the result of the 5-pixel convolution for the high-level filter.
The low-level filter is constructed by application of a threshold on the difference of convolutions with the small and large kernel.
Detections in the areas affected by incomplete sampling due to image motions are discarded in both filters.
Several artifacts caused by hits of cosmic rays are also removed.
Finally, a detection of the low threshold is accepted if there are detections reaching the high threshold at three or more different times.

\clearpage
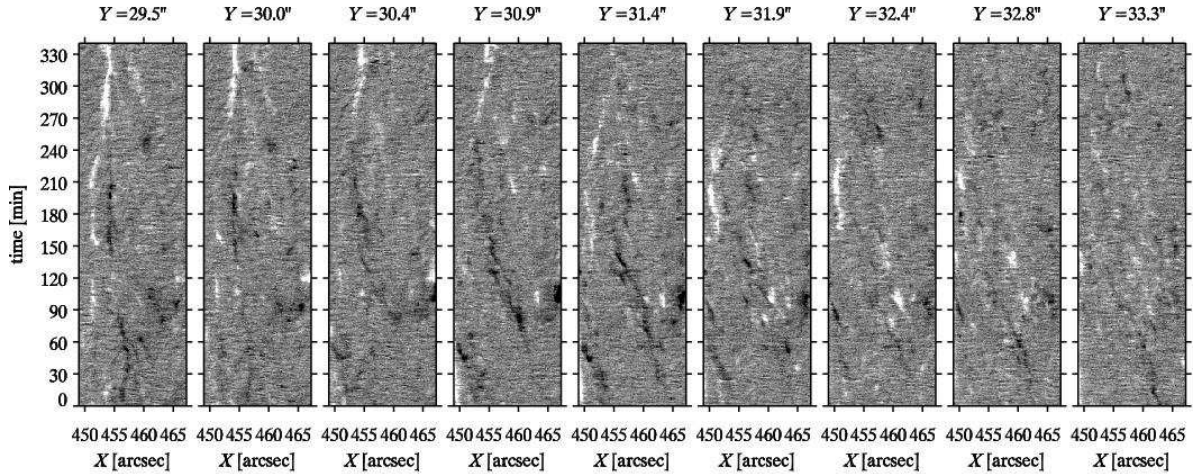
\figuretwo{f2}{fig:slicesample1}{A sample set of $X$--$t$ slices of the magnetograms at successive $Y$ positions.
The region is indicated in Fig.~\ref{fig:sample}.
Each $X$--$t$ slice shows the density of flux along the line of sight averaged over 3 pixels ($0.48\arcsec$) in $Y$ at the indicated $Y$ position.
The step size in $Y$ is also 3 pixels.
The flux density is scaled between $-200$ (black) and $+200~\mxsqcm$ (white).}

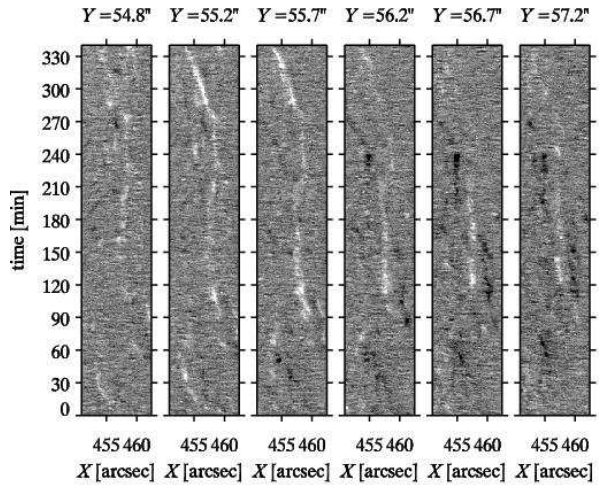
\figureone{f3}{fig:slicesample2}{A second set of sample $X$--$t$ slices of the magnetograms at successive $Y$ positions.
The format is identical to the format in Fig.~\ref{fig:slicesample1}.}
\clearpage

The algorithm identifies 11579 magnetic elements with positive flux, and 5226 with negative flux.
Visual inspection of the result shows that this procedure rejects noise adequately and often captures weak fields.
Obviously, structures with very short lifetimes, i.e., less than three time steps, are discarded by our algorithm.
We believe this is no serious issue for the present analysis, because we are primarily interested in longer-lived structures.

In this paper, we focus on small-scale concentrations of vertical field in internetwork areas, that we will refer to as ``internetwork magnetic elements'' or IMEs.
We discard elements when they overlap with network areas.

\section{Analysis, Results, and Discussion}\label{sec:analysis}

\subsection{Visual inspection}\label{sec:visual}

We searched for patches of IMEs in the sequence of magnetograms.
To this end, we employed a ``cube slicer'' dissecting the data cube into $X$--$Y$ and $X$--$t$ slices, similar to the procedure by 
	\citet{2005A&A...441.1183D}, 
who inferred the existence of long-lived ``magnetic patches'' in quiet sun internetwork areas from recurrent bright points in G-band and \CaIIH\ sequences.
We found that while many long-lived structures of one or more IMEs exist, they do not typically show strong location memory on timescales of several hours.
Some remain stationary for long periods, while others may migrate significant distances in a few hours.
MEs often interact with other MEs while migrating, making the identification of well-defined structures difficult.
In addition, many short-lived concentrations of flux appear frequently in the vicinity of longer-lived structures, but with unclear association to a particular structure.

We cannot combine IMEs into ``patches'' following the procedure of
	\citet{2005A&A...441.1183D}. 
Their proxy-magnetometry likely misses many weak and short-lived structures.
In the present data, the density of IMEs, especially of short-lived flux concentrations, and the many interactions of long-lived structures inhibits successful grouping of elements into well-defined patches.
The parameters that govern grouping of elements can be tuned to yield either many short-lived patches consisting of a single IME, or a small number of long-lived patches containing the vast majority of IMEs.

Figure~\ref{fig:slicesample1} imitates ``cube slicing''
	\citep[cf.\ Sect.~2 of][]{2005A&A...441.1183D}
on one of the regions of interest indicated in Fig.~\ref{fig:sample}.
It shows a sequence of $X$--$t$ cutouts at progressive $Y$ locations.
The left edge of the region is in the magnetic network (see Fig.~\ref{fig:sample}).
Many short-lived concentrations of flux with both positive and negative polarity are visible throughout the region.
In addition, several elements with negative polarity show up as trails toward network in the slices from $Y=30.9\arcsec$ to $Y=33.3\arcsec$.
They start around $X\approx465\arcsec$ at $t=0$.
At $t\approx150~\min$, their motion in $X$ slows as they reach the stationary element around $X\approx455\arcsec$ and $Y\approx30.0\arcsec$.
True cube slicing shows that while the IME is continuously present, it sometimes drops below the detection limit and reappears some time later.

Figure~\ref{fig:slicesample2} shows a different region in the same format as Fig.~\ref{fig:slicesample1}.
In this case, an IME of positive polarity appears around $X\approx460\arcsec$ at $t\approx90~\min$, and remains visible until the end of the sequence.
It remains stationary at first, then migrates quickly in $X$ starting from $t\approx270~\min$.
Again, many short-lived concentrations of flux of both polarities are visible throughout the region.
Some short-lived elements appear to be part of larger structures.
In the $Y=56.7\arcsec$ panel, for instance, the concentration of negative polarity around $X\approx453\arcsec$ and $t\approx270~\min$ seems connected to the stable element around $X\approx455\arcsec$ and $t\approx230~\min$ by a faint trail of negative flux best visible in the $Y=56.2\arcsec$ panel.
Similarly, intermittent concentrations of positive flux around $X\approx454\arcsec$ and $t\approx30~\min$ at $Y=55.2\arcsec$ may be connected to the long-lived structure around $X\approx459\arcsec$ and $Y=55.7\arcsec$, as may elements around $X\approx455\arcsec$ and $t\approx240~\min$ at $Y=54.8\arcsec$.

Figure~\ref{fig:samplegm} shows a sample mask of magnetic elements.
It corresponds to the frames shown in Fig.~\ref{fig:sample}.
IMEs are not spread homogeneously over internetwork, but rather appear to outline cells on scales of several Mm.

\clearpage
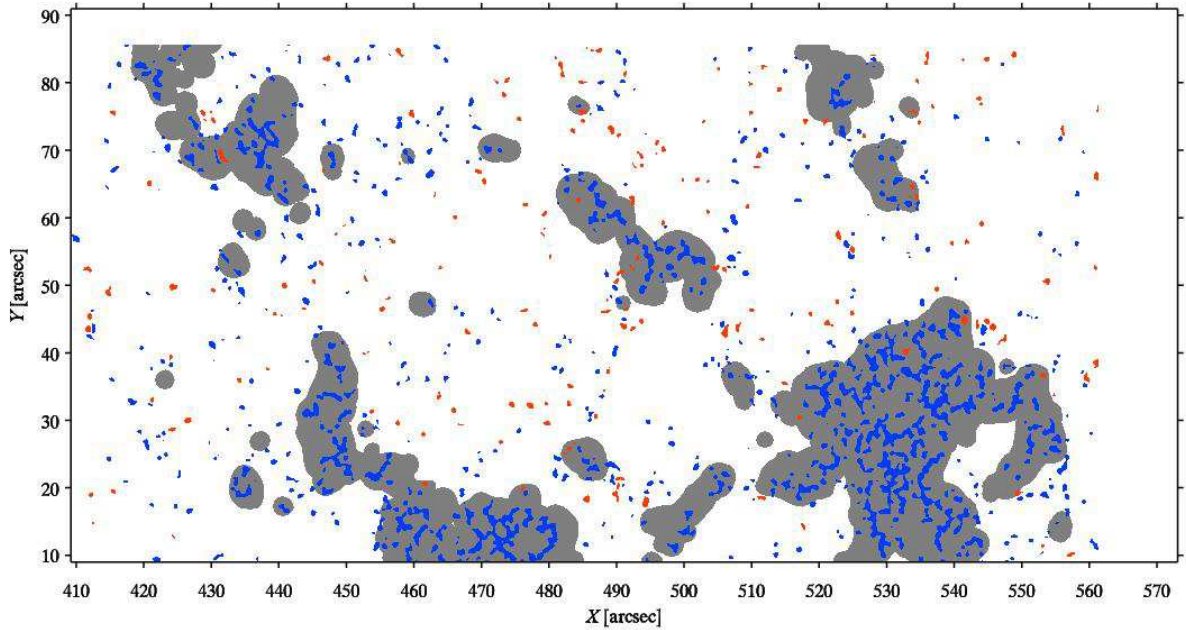
\figuretwo{f4}{fig:samplegm}{Sample mask of magnetic elements corresponding to the frames shown in Fig.~\ref{fig:sample}.
Elements with positive flux are blue, those with negative flux red.
Network is shaded gray.}
\clearpage

\subsection{Comparison with bright points}\label{sec:comparebp}

We search for bright points in the G-band and \CaIIH\ image sequences at those locations where our algorithm detects an IME\@.
A random IME at a random time during the sequence is selected, and the magnetogram at that location is manually compared to the G-band and \CaIIH\ intensities.
Fragmented IMEs are treated as if they were multiple IMEs.
IMEs that have no clear visually identifiable flux are discarded.
Such IMEs typically have strong Stokes V signal in the preceding or following frame.
They are detected because the identification algorithm averages over three time steps.
We so inspect $400$ IMEs.
Figure~\ref{fig:globsamples} shows three samples.
A bright point is identified in the G-band intensity in $88$ cases ($22\%$), and in the \CaIIH\ intensity in $73$ cases ($18\%$).
$67$ cases ($17\%$) exhibit a bright point in both the G-band and \CaIIH\ intensities.
Due to the small number of samples, the uncertainty on these measurements amounts to $5\%$.
While the difference between G band and \CaIIH\ is statistically insignificant, the numbers do agree with the impression that bright points are more easily visible in the G-band than in the \CaIIH\ line.
Weak bright points are drowned in the background of reversed granulation in \CaIIH, while G-band bright points appear in the dark intergranular lanes.

These findings agree with ground-based results of
	\citet{2001ApJ...553..449B} 
and
	\citet{2007A&A...472..911I}, 
who found that magnetic field is a necessary, but not a sufficient condition for the formation of bright points in G-band or \CaIIK\ intensity.
The visual inspection reveals no evident relationship between morphology of the IME and the existence of a bright point in the G-band or \CaIIH\ intensity.
Some IMEs appear small and weak, either because the field is weak, or because it is oriented away from the line of sight, yet have clear bright points (top row of Fig.~\ref{fig:globsamples}), whereas other strong elements do not show any brightening (bottom row of Fig.~\ref{fig:globsamples}).

Most IMEs that have an associated \CaIIH\ bright point also have a bright point in G-band intensity.
This suggests that the mechanism of brightness enhancement is similar.
G-band bright points are formed by weakening of molecular CH lines
	\citep{2006ApJ...639..525U}, 
as a result of partial evacuation of the magnetic element
	\citep{1976SoPh...50..269S, 
	2004ApJ...607L..59K, 
	2004ApJ...610L.137C}. 
In the wings of \CaIIHK, the magnetic element is cooler at equal geometrical height, but hotter at equal optical depth
	\citep{2005A&A...437.1069S}. 
Since the \CaIIH\ filter of \emph{Hinode} is $0.3~\nm$ wide
	\citep{Tsuneta2007},
the bulk of the emission is formed in the wings of the line.
The root cause of brightness excess in both the G-band and \CaIIH\ filtergrams used here is the fluxtube Wilson depression.
The chromospheric contribution of the core of the \CaIIH\ line will become more important as narrower filters are employed, and it is likely that correspondence between bright points in the G-band and \CaIIH\ intensities will be reduced.

\clearpage
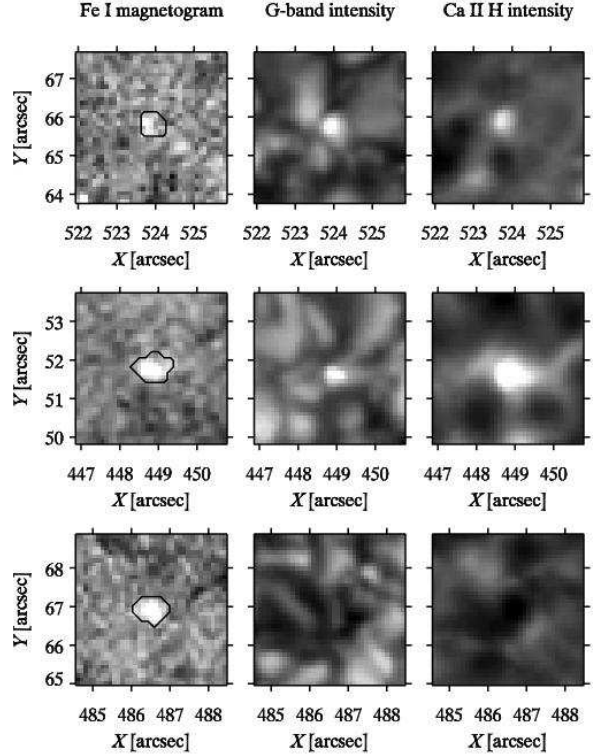
\figureone{f5}{fig:globsamples}{Sample IMEs with (top two rows) and without (bottom row) associated bright points.
The IME mask is indicated by the contour in the magnetogram.
The magnetogram is scaled between $-200$ (black) and $+200~\mxsqcm$ (white).
The top row shows an IME that has little apparent flux, yet still produces a clear bright point in both the G-band and \CaIIH\ intensity.
The IME in the bottom row is similar to the IME in the middle row, but does not have associated bright points in either the G-band or \CaIIH\ intensity.}
\clearpage

\subsection{Lifetime}\label{sec:lifetime}

We next compute lifetimes for each IME\@.
Figure~\ref{fig:hlt} shows the histograms of IME lifetimes, for positive and negative polarity separately.
Elements have an average measured lifetime of about $10~\min$.
It should be noted that accurately determining the lifetime is difficult and prone to errors
  \cite[see Sect.~3 of][]{2007ApJ...666..576D}. 

The average lifetime resembles results of
	\citet{1998ApJ...495..973B}, 
who measured a mean lifetime of $9.3~\min$ for bright points in network areas.
However, it is significantly longer than those measured for internetwork bright points in G-band ($3.5~\min$) and \CaIIH\ images ($4.3~\min$) by
	\citet{2005A&A...441.1183D}. 
Based on positions of their bright points, those authors argued that bright points map positions of long-lived fields that exist before a bright point appears and persist after it disappears.

The present analysis allows us to track magnetic elements that no longer have an associated bright point, yet we do not find lifetimes of several hours, predicted by
	\citet{2005A&A...441.1183D} 
based on statistical analysis of groups of bright points in a 1-hour sequence.
However, it is clear that many elements live longer than our detection algorithm is able to track them from visual inspection of the magnetograms (cf.\ Sect.~\ref{sec:visual}).
The lifetime computed here is a measure of how long flux remains concentrated enough for our algorithm to track it, rather than an estimate of the lifetime of flux itself.

\clearpage
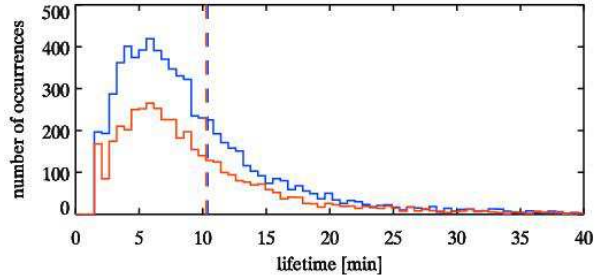
\figureone{f6}{fig:hlt}{Histograms of the lifetime of IMEs.
Blue: positive polarity.
Red: negative polarity.
The averages are indicated by the vertical dashed lines.}
\clearpage

\subsection{Proper motions}\label{sec:velocities}

We measure the positions of IMEs by computing their center of mass, using the flux density from the magnetograms for weighing.
First, elements are discarded if they reach the edges of the area affected by incomplete sampling due to image motions (dotted white lines in the bottom left Fig.~\ref{fig:sample}).
Splitting or merging elements are then cut up into separate detections by wiping out the IME in the frame after the merger, or in the frame before the split.
Finally, we identify elements whose mask changes between consecutive frames by over five times the minimum mask size in those frames.
This can be caused by the appearance of a new element close to an existing element, so that it is identified as the existing element.
For those cases, the IME in the second frame is erased.
For the remaining detections, the center of mass of the mask is computed, while weighing with the magnetograms smoothed over three frames in time.
We so find 136\,271 and 76\,211 positions for IMEs with positive and negative flux, respectively.

Hereafter we use the term `velocity' for the inferred proper motions of IMEs.
A velocity can only be computed if a magnetic element has a well-defined position in two consecutive frames.
This reduces the number of usable positions to 112\,871 measurements for IMEs with positive flux and 62\,748 for those with negative flux.
The top two panels in Fig.~\ref{fig:hv} show the distribution of proper motions in $X$ and $Y$.
They are nearly Gaussian around the origin with a slight overdensity at velocities close to zero and in the far wings.
Disregarding polarity and direction, a Gaussian fit gives an average of $0.02\pm0.10~\kms$ and an rms of $1.57\pm0.08~\kms$.

The rms of the proper motions is nearly a factor of three higher than that measured by
	\citet{1998ApJ...509..435V}, 
who found an rms of $0.54~\kms$ using a cork tracking technique.
However, the present analysis is in good agreement with the measurement of an rms of $1.31~\kms$ from
	\citet{2003ApJ...587..458N}. 
Those authors attribute the higher velocities to the selection of dynamic isolated bright points, rather than stable network elements.
Since IMEs are typically also isolated, the results presented here are affected by the same selection effect.

The histogram of the speed shows the expected Rayleigh distribution, though it has a noticeably longer tail toward higher speeds.
Finally, there is no evidence of a correlation between the velocity in $X$ and the velocity in $Y$ in the scatter plot (Fig.~\ref{fig:hv}, lower right panel).

In total, there are 2426 measurements ($1.4\%$) of proper motions larger than $8~\kms$.
Investigation shows that these high velocities are caused by inaccuracies in the determination of the center of mass of weak elements as a result of noise in the magnetograms.

\clearpage
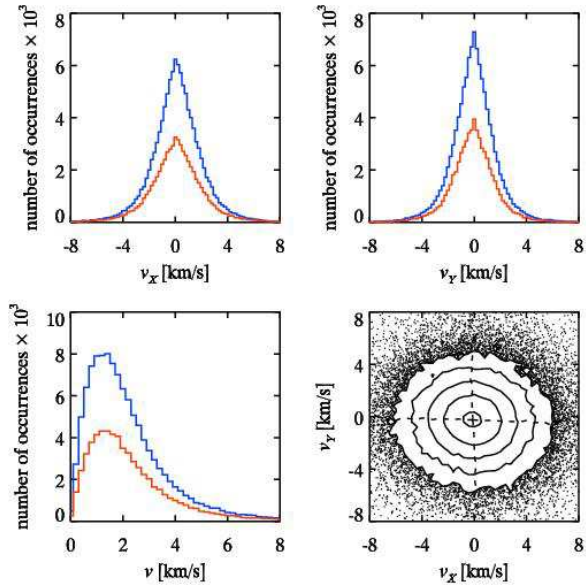
\figureone{f7}{fig:hv}{Histograms of the velocity of IMEs in $X$ (top left panel) and $Y$ (top right panel), and of the speed of IMEs (bottom left panel).
Blue: positive polarity.
Red: negative polarity.
The bottom right panel shows velocity in $X$ vs velocity in $Y$, combining both polarities.
Dashed lines indicate the first moments.}

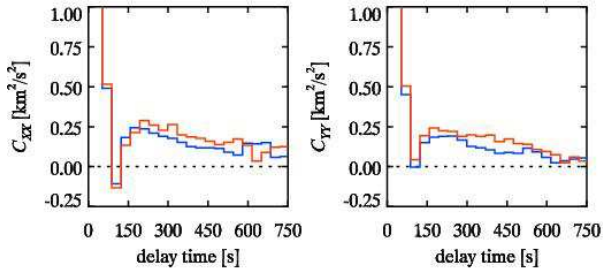
\figureone{f8}{fig:acv}{Autocorrelations of the velocity of IMEs in $X$ (left panel) and $Y$ (right panel).
Blue: positive polarity.
Red: negative polarity.}
\clearpage

Figure~\ref{fig:acv} shows the autocorrelation of the velocities in $X$ and $Y$.
Errors in measurement of position propagate into the autocorrelation of the velocity
	\citep[cf.][]{2003ApJ...587..458N}. 
The first four measurements around $\Delta t=0$ are influenced by the present analysis because it averages three frames in time.
While this precludes determining the correlation time of IME velocities, the autocorrelations are consistently positive for delays up to $600~\s$, in agreement with results from bright point tracking by
	\citet{2003ApJ...587..458N}. 

\subsection{Direction}\label{sec:direction}

We speculate that the low, yet positive, autocorrelation at long delays is caused by a consistent, slow drift of the IMEs toward network.
The component of velocity toward the nearest network boundary was computed over one time step, corresponding to $35~\s$.
Elements that are closer to the area affected by image motion than to the nearest network pixel are discarded.
The results are shown in Fig.~\ref{fig:hvtonw}.
IMEs display a broad velocity distribution, with a slight bias toward the network boundary.
Using a Gaussian fit, we measure a bias toward the network boundary of $0.152\pm0.006$ and $0.218\pm0.009~\kms$, and rms velocities of $1.746\pm0.005$ and $1.840\pm0.007~\kms$, for IMEs with positive and negative flux, respectively.

Over 17~time steps ($597~\s$), close to a granular turn-over timescale, the average velocity changes only little, to become $0.104\pm0.006$ and $0.227\pm0.010~\kms$ for IMEs with positive and negative flux, respectively.
The rms is reduced significantly to respectively $0.608\pm0.005$ and $0.703\pm0.008~\kms$.

The average velocity toward the network boundary is higher for IMEs with negative flux.
The difference between the polarities is statistically significant up to velocity measurement over 35~time steps ($1230~\s$), and tends to grow as the number of time steps increases.
The number of measurements drops below acceptable levels for longer delays.
Perhaps the higher mobility of elements with negative flux is a result of an attraction to the network, which consists predominantly of positive flux.
Alternatively, it may be a result of higher mobility of IMEs of negative polarity due to smaller size (median size of 14~pixels in the detection mask) compared to IMEs of positive polarity (median size of 17~pixels).

\clearpage
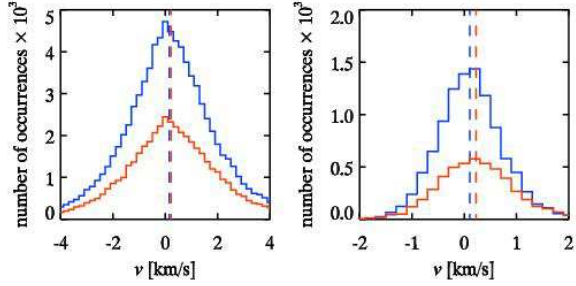
\figureone{f9}{fig:hvtonw}{Histograms of the velocity of IMEs toward the network boundary measured over one time step ($35~\s$; left panel) and 17~time steps ($597~\s$; right panel).
Blue: positive polarity.
Red: negative polarity.
The best-fit bias velocities are indicated by the vertical dashed lines.
Both polarities show a broad distribution with a slight bias toward velocities in the direction of the nearest network pixel.
The width of the distribution diminishes with increasing timescales, while the bias does not change noticeably.}
\clearpage

The left panel of Fig.~\ref{fig:cfa} shows the distribution of the directions of velocity over one time step.
It shows a preference for velocities with directions around $0^\circ$ and $180^\circ$, i.e., along the $X$~axis.
Correspondingly, the peak of the velocity distribution is higher in $Y$ than it is in $X$ (top panels in Fig.~\ref{fig:hv}), and the scatterplot of $v_Y$ against $v_X$ is somewhat oval, indicating there are more IMEs moving at higher speed in the $X$.
In the classical view that IMEs are advected by gas motion, this preference in the direction of propagation is should be caused by supergranular flows.
Figure~\ref{fig:sample} shows that the supergranular cell is somewhat elongated in $X$, suggesting that the flow is indeed stronger in that direction.

We also compute the centrifugal acceleration $v\,\mathrm{d}\phi/\mathrm{d}t$, where $\phi$ is the direction of the IME motion.
This is the relevant quantity when considering generation of transverse waves in fluxtubes.
Measurements with velocities below $1~\kms$ are discarded, because the errors in the determination of the angle increases with decreasing velocity.
The histogram of the centrifugal acceleration is shown in the right panel of Fig.~\ref{fig:cfa}.
The distribution is Gaussian with an average of $0.045\pm0.013$ and an rms of $4.559\pm0.011~\mathrm{deg}\;\mathrm{km}/\mathrm{s}^2$.
	\citet{2003ApJ...587..458N} 
found similar results, however, a detailed comparison is hampered by their low number of measurements.

\clearpage
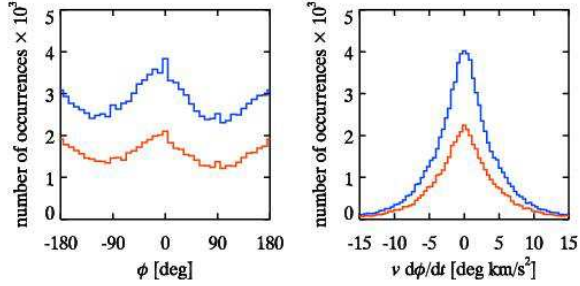
\figureone{f10}{fig:cfa}{Histograms of the direction of motion (left panel) and the centrifugal acceleration (right panel) of IMEs.
Blue: positive polarity.
Red: negative polarity.}
\clearpage

\subsection{Emerging bipoles}\label{sec:emerging}

\clearpage
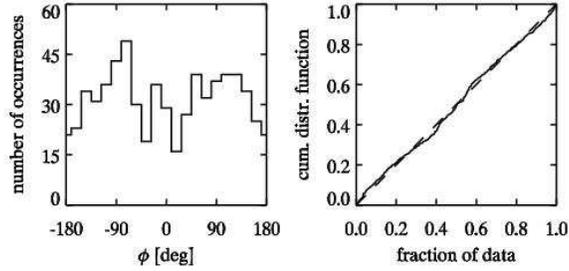
\figureone{f11}{fig:epp}{Histogram of the distribution of angles between emerging bipoles (left panel), and the probability-probability plot of the distribution against the hypothesis of uniform distribution (right panel).
The dashed line represents the cumulative distribution function of a uniform angular distribution.}
\clearpage

We search for emerging bipoles by pairing magnetic elements of opposite polarity that appear close in time and space.
A pair of magnetic elements of opposite polarity is flagged if the elements appear within a radius of $1.4~\Mm$ and no more than four time steps ($140~\mathrm{s}$) apart.
There are 639 pairs that satisfy these conditions.
For these pairs we next compute the angle of the line connecting the center of the negative element with the center of the positive one.

Figure~\ref{fig:epp} shows the directional distribution and probability-probability (``p-p'') plot of the angle under the assumption that the angular distribution is uniform.
A p-p plot can be used to see if a set of data follows a given distribution.
It is constructed by plotting the cumulative distribution function $F(x_i)$ against $(i-\frac12)/n$, where $x_i$ is the $i$th data point, ordered from smallest to largest, and $i=1,2,\dots,n$.
Here, we have taken a uniform distribution of angles $F(x_i)=(x_i+180)/360$.
If the data is perfectly uniformly distributed, we have $x_i=360\,(i-\frac12)/n-180$.
Substitution yields $F(x_i)=(i-\frac12)/n$, so that a linear p-p plot results.
Since the p-p plot in Fig.~\ref{fig:epp} is approximately linear, there does not appear to be preference in orientation of the emerging magnetic elements.
We conclude that there is no evidence for Joy's law at arcsecond scales from these data, in agreement with results of
	\cite{2007AAS...210.9213L}. 

Pairs that are not actual emerging bipoles have random angles, and therefore only add a statistically uniform background that does not contribute signal to the directional distribution or the p-p plot in Fig.~\ref{fig:epp}.
However, the number of accidental associations may be large, so that any signal from real emerging bipoles is drowned in noise.
A search for emerging flux using \emph{Hinode}'s spectropolarimeter, such as the one by
	\cite{2007ApJ...666L.137C} 
would not suffer from potentially large numbers of false positives, even if automated, and would therefore give more trustworthy results.

\section{Summary and Conclusion}\label{sec:conclusion}

We have analyzed the dynamics of IMEs using a sequence of magnetograms.
Visual inspection of the data shows the existence of many long-lived magnetic elements that have frequent interactions with other elements during their lifetime.
We find that they may migrate over distances of $\sim7~\Mm$ in periods of several hours.
Their interactions, migration, and the many short-lived concentrations of flux that appear in their vicinity make it cumbersome to uniquely identify an IME, or a set of IMEs.
The IMEs sometimes drop below the detection level in our data, but commonly reappear some time later.
IMEs appear to outline cells on scales of several Mm.

A manual inspection of IME locations in the G-band and \CaIIH\ filtergrams shows that only about a fifth of the IMEs have associated bright points.
Visual inspection reveals no obvious correlation between IME morphology and the existence of bright points.
There is a substantial correlation between the existence of a bright point in the G-band and \CaIIH\ intensities.
Bright points are formed in the G band through weakening of molecular CH lines as a result of partial evacuation of the magnetic element, while they are caused in \CaIIH\ by influx of radiation from the hot walls of the Wilson depression.
We therefore attribute the similarity in the appearance of bright points in these passbands to their common origin.

We identity magnetic elements using a hysteresis-based algorithm that is able to track IMEs for about $10~\min$ on average.
This is much shorter than the lifetime of several hours predicted by
	\citet{2005A&A...441.1183D}. 
However, visual inspection shows that many elements intermittently drop below the detection limit of our algorithm, shortening their measured lifetime.

IMEs exhibit proper motions that resemble a Gaussian distribution with a slight overdensity of velocities near the origin and in the far wings.
We measure an rms velocity of $1.57\pm0.08~\kms$.

The IMEs show a slight bias of about $0.2~\kms$ for velocities toward the nearest network boundary.
This bias persists up to timescales of at least $1200~\s$.
It is the likely cause of weak positive velocity autocorrelation at long delay times.
In the data analyzed here, IMEs of negative polarity show a statistically significant higher drift to the nearest network boundary.
It is tempting to assume that this difference is somehow related to the dominant positive polarity of the network in the field of view.
Alternatively, elements of negative polarity may be more mobile compared to elements of positive polarity because of their smaller size.

We find a slight preference for velocities in $X$~direction.
Elongation in $X$ of the main supergranular cell in the field of view suggests that supergranular flows are stronger in that direction.
The observed preference in direction is thus in agreement with the classical picture that IMEs are advected by supergranular flows.

IMEs experience centrifugal accelerations that obey a Gaussian distribution with an average of $0.045\pm0.013$ and an rms of $4.559\pm0.011~\mathrm{deg}\;\mathrm{km}/\mathrm{s}^2$, in agreement with results of
	\citet{2003ApJ...587..458N}. 

We search for emerging bipoles by pairing elements of opposite polarity that appear nearby each other in space and time.
There is no detectable preference in the orientation of a pair.
We conclude, therefore, that there is no evidence to support Joy's law at arcsecond scales from these data.

\acknowledgments{\emph{Hinode} is a Japanese mission developed and launched by ISAS/JAXA, collaborating with NAOJ as a domestic partner, NASA and STFC (UK) as international partners.
Scientific operation of the \emph{Hinode} mission is conducted by the \emph{Hinode} science team organized at ISAS/JAXA.
This team mainly consists of scientists from institutes in the partner countries.
Support for the post-launch operation is provided by JAXA and NAOJ (Japan), STFC (UK), NASA, ESA, and NSC (Norway).
AdW thanks P.~Judge, S.~McIntosh, R.J.~Rutten, and J.~Trujillo Bueno for discussions.}

\let\clearpage\tmpclearpage

\end{document}